\documentclass[12pt]{article}
\usepackage{epsf}
\usepackage{epsfig}
\usepackage{graphicx}

\begin{document}

\begin{center} 
{\bf \Large Observation of Z Decays to b Quark Pairs}\\
{\bf \Large at the Tevatron Collider} \\  
\mbox{} \\
\mbox{} \\            
Tommaso Dorigo \\
(Padova University and INFN)\\
{\em for the CDF Collaboration}\\
\mbox{} \\
\end{center}
      
\begin{center}
\textbf{Abstract}
\end{center}

\noindent
\begin{center}
\begin{minipage}{13cm}
{\footnotesize A search for Z boson decays to pairs
of $b$-quark jets has been performed in the full dataset collected with
the CDF detector at the Tevatron $p\bar{p}$ collider.
After the selection of a pure sample of $b\bar{b}$ events by 
means of the identification of secondary vertices from $b$-quark decays, 
we have used two kinematic variables to further 
discriminate the electroweak $b\bar{b}$
production from QCD processes, and sought evidence
for the Z decay in the dijet invariant mass distribution. An
absolute background prediction
allows the extraction of an excess of events 
inconsistent with the background predictions by $3.23 \sigma$ but in 
good agreement with the amount and characteristics of the expected signal.
We then fit the mass distribution with an unbinned likelihood technique,
and obtain a $Z \to b\bar{b}$ signal amounting to $91\pm 30 \pm 19$ events. } 
\end{minipage}
\end{center} 
\vspace {1.5cm}

\section {Introduction}

\noindent 
Z decays to $b$-quark pairs are not exactly a unknown piece of Physics. 
Since 1992 the LEP experiments have detected several 
millions of them, and more have come from the polarized beams of the SLC. 
The process is thus very well understood; the Z is one of the best known
particles, and there can be no surprise in the thereabouts. 
At a proton-antiproton collider this particular process has never been 
seen before, though. The UA2 collaboration published in 1987 an analysis 
of jet data where they could spot the combined signal of W and Z 
decays to dijet pairs\footnote
   {{\em Phys. Lett.} 186B (1987), 452. The search
   was later updated with a larger data sample which produced a signal 
   of $5,367\pm958$ events: {\em Z. Phys.} C49 (1991), 17.}, 
but the decay of the Z to $b$ quarks was not separated from the other 
hadronic decays. 

   The Z decay to $b$ quarks is
the closest observable process to the expected decay of the Higgs boson, and 
---in view of the chances of a Higgs discovery in run 2 at the Tevatron--- the 
understanding of the process, the knowledge of the expected mass resolution for 
a reconstructed decay, and the confidence with the 
kinematical tools that may help extracting it, are topics worth careful investigation.  
Moreover, a clean Z peak in a dijet mass distribution is useful for
jet calibration, a relevant issue for the top quark mass measurement.

   In the following we present a search for the $Z \to b \bar{b}$ process in    
$L=110 pb^{-1}$ of $p \bar{p}$ collisions collected by CDF during 
the years 1992-1995. We will show that the signal 
can be extracted by means of a very stringent selection 
that allows a reduction of the background by four orders of 
magnitude, while retaining in the final dataset a handful of Z events. 
 
   This paper is organized as follows: in Section 2 we describe the CDF detector, 
the datasets used in the analysis, and the selection that leads to our final sample; 
in Section 3 the counting experiment is described, and the results are discussed; 
Section 4 deals with the unbinned likelihood fits to the mass distribution. 
In Section 5 we present our conclusions.

\section{ Data Selection }

\noindent
The CDF detector has been described in detail elsewhere\footnote{
      F.Abe {\em et al.}, Nucl. Instr. Meth. Phys. Res. A {\bf 271} (1988),
      387, and references therein.}. 
We only mention briefly here those detector components most relevant to this analysis.
The silicon vertex detector (SVX) and the central tracking chamber (CTC) are 
immersed in a $1.4 \, T$ axial field and provide
the tracking and momentum measurement of charged particles. The SVX consists of 
four layers of silicon microstrip detectors and provides spatial measurement
in the $r-\phi$ plane\footnote {In CDF the positive z axis lies along the proton
   direction, r is the radius from the z axis, $\theta$ is the polar angle, and $\phi$
   is the azimuthal angle.}                            
with a resolution of $15 \,\mu m$\footnote{
   D.Amidei {\em et al.}, Nucl. Instr. Meth. Phys. Res. A {\bf 350} (1994), 73;
   S.Tkaczyk, Nucl. Instr. Meth. Phys. Res. A {\bf 368} (1995), 179, and 
   references therein.}. 
The CTC is a cylindrical drift chamber
containing 84 layers grouped into 9 alternating superlayers of axial and stereo
wires. The calorimeters, divided into electromagnetic and hadronic components,
cover the pseudorapidity range $|\eta|<4.2$ and provide the jet energy
measurement; they are segmented into  projective towers subtending
about $0.1 \times 15^{o}$ in $\eta - \phi$ space. Central muon candidates are 
identified in two sets of muon chambers, located outside the calorimeters, 
as stubs extrapolating to charged tracks inside the solenoid. The data were collected 
by a three-level trigger system; the first two levels are provided by hardware 
modules, the third consists in software algorithms optimized for speed.

   At the Tevatron the Z production cross section has been measured both in the 
$e^+ e^-$ and in the $\mu^+ \mu^-$ final states, and found to be 
$\sigma_Z \times BR(Z \rightarrow e^+e^-) = 0.235 \, \pm \, 0.003 \, (stat.) \, \pm  \,
0.005 \, (syst.) \, \pm \, 0.020 \, (lum.) \, nb$ and $\sigma_Z \times 
BR(Z \rightarrow \mu^+ \mu^-) = 0.202 \, \pm \, 0.016 \, (stat.) \, \pm \, 0.020$  
$(syst.)$  $\pm \, 0.017 \, (lum.) \, nb$\footnote{
     See for instance P.Quintas, {\em Proc. XI Symposium on Hadron Collider Physics}, 
     World Scientific, Singapore 1996.}. 
By scaling the $e^+ e^-$ figures up to the b quark branching fraction\footnote{
   We can use for that purpose the world average branching fractions 
   $\Gamma_{Z \to b \bar{b}}/ \Gamma_{Z \to hadrons} = 0.2212 \pm 0.0019$ and
   $\Gamma_{Z \to hadrons}  / \Gamma_{Z \to e^+ e^-} = 20.77  \pm 0.08$
   (PDG 1996). }
one expects that about 110,000 Z decays to $b$-quark pairs have taken place at 
CDF during run 1, the cross section 
times branching ratio for the process resulting in 
$1.080 \pm 0.029 (stat. \oplus syst.) \pm 0.092 (lum.)$ nanobarns. 

   The natural trigger for a $Z \to b \bar{b}$ signal would be a low energy
jet trigger. However, the rate of jet production processes is too high 
to allow for a collection of all these events: the jet triggers are 
therefore {\em prescaled}, such that only a limited integrated luminosity 
is collected by them. The best dataset for a search of $Z \to b \bar{b}$ decays
is instead the one collected by a trigger requiring the presence of a 
clean muon candidate: the 
$b$-quark jets produced in Z decays contain lots of low $P_T$\ muons, 
originated in B hadron decays, sequential charmed hadron decays, 
or other processes in smaller amounts. The starting point of the search
was therefore the sample of 5.5 million events featuring a 
$P_T>7.5 \, GeV/c$ central muon candidate, corresponding to an integrated 
luminosity of $L=103 \pm 7 \, pb^{-1}$.

\subsection { The Datasets }

\noindent
The initial dataset was cleaned up by requiring the muon candidate
be also identified as a good muon by a standard offline filter. 
Jets were reconstructed by a fixed $R=0.7$ cone clustering algorithm\footnote {
    F.Abe {\em et alii}, Phys. Rev. D {\bf 45} (1992), 1448. },      
and two jets in each event were required to contain charged tracks forming a 
well identified secondary vertex (hereafter ``{\em SVX tag}'', from the name of 
the silicon vertex detector). 
This selection reduced drastically the dataset, being
satisfied by only 5,479 events. The selection discarded the 
non-heavy flavour component of the QCD background, and increased the
signal/noise ratio by about two orders of magnitude.     

   We used the PYTHIA Monte Carlo\footnote {
      H.Bengtsson and T.J.Sjostrand, Comput. Phys. Commun. {\bf 46} (1987), 43. }   
to generate 1.7M $Z \to b\bar{b}$ decays 
that were subjected to a detector simulation, filtered by a trigger 
simulation, and passed through the same offline selection used for the real data. 
Using the predicted cross section for the searched process, the   
number of $Z \to b\bar{b}$ decays in the double SVX tagged dataset is expected 
to amount to $124\pm14$ events.

   As will be explained in the following, we have used real Z decays to 
electron-positron pairs collected during run 1 to better understand the 
behavior of initial state radiation in the Z production. 
These events were selected from high $E_T$ 
electron triggers by requiring the presence of a good quality 
central electron plus another electron candidate passing looser cuts. 
The dataset consists in more than six thousand events,
and can be useful for kinematical studies, particularly when the 
variables one is interested in are difficult to model with Monte Carlo
programs. 

   To study the contamination of other boson signals in our dataset we also
generated 500,000 W boson decays to $c \bar{s}$ pairs and 500,000 Z decays
to $c$-quark pairs with PYTHIA 5.7: in fact, these processes may give a 
contribution to the muon dataset, due to the presence of charm quarks in the
final state. These events were subjected to the same treatment described for the
$Z \to b\bar{b}$ data. Their contamination to the double SVX tags dataset
was found to be totally negligible.

 \begin{table}[htb]
\centering
\begin{tabular}{||l||c|c|c|c||}
\hline
\hline
Sample                 &   Run 1     & $Z\to b\bar{b}$  & Eff.      & Events in        \cr
                       &             &  PYTHIA          &           & 103 $pb^{-1}$    \cr
\hline
                       &             &                  &           &                  \cr
Initial                &             & 1,673,000        &           &$(110\pm12)10^3$  \cr
Trigger                &  5,414,755  &                  &           &                  \cr
2 SVX tags             &      5,479  &     1,867        &  21.3\%   & $ 124\pm 14 $    \cr
$\Delta \Phi_{12}>3$   &      1,684  &     1,368        &  73.2\%   & $  91\pm 11 $    \cr
$\Sigma_3 E_T<10   $   &        588  &                  & (50.0\%)  & $ (45\pm 21)$    \cr
$$ & $$ & $$ & $$ & $$  \cr
\hline
\hline
\end{tabular}
\caption{ {\em Number of events selected by each of the cuts described in the
          text for the experimental data and for the} $Z \to b\bar{b}$ {\em MC data,
          efficiency of each cut for the Z signal, and expected Z events in the 
          total dataset. The signal 
          size and the efficiency of the last cut 
          (}$\Sigma_3 E_T<10 \, GeV${\em ) are estimated using} 
          $Z \to e^+ e^-$ {\em data. } 
          \label{table:cuts}}
\end {table}

\subsection { Kinematical Tools }

\noindent
Even after the very restrictive selection of events with two SVX tags, 
the signal/noise ratio is too small to allow for a clear identification of 
the $Z \to b \bar{b}$  
signal in the dijet mass distribution: other tools are needed 
to increase the discrimination of the electroweak production of $b$-quark jets 
from the strong interaction. 

   From a theoretical standpoint, one expects the two production processes 
to be different in many ways. The two initial state partons of the Z 
production have to be a quark-antiquark pair, and the process is time-like. 
On the contrary, in the QCD creation of a $b\bar{b}$ pair both a 
quark-antiquark and a gluon-gluon pair can give rise to a time-like direct 
production process, and also space-like diagrams may contribute. 
Many of the QCD processes are expected to result in a pair of outgoing 
partons with a flatter pseudorapidity 
distribution than those from the searched Z decay; but the double SVX 
tagging and the
requirement of a central muon candidate in one of the two jets result in 
pseudorapidity distributions that are already very constrained and 
well peaked at zero, due to the acceptance of the SVX and the muon chambers. 

   If one examines the color structure of the diagrams one however notices 
a marked difference between $Z \rightarrow b\bar{b}$ and 
$g \rightarrow b\bar{b}$. In the QCD processes 
there is a color connection between the initial and the final state 
absent in the Z production. Furthermore, while both the initial and the 
final state of the Z production are in a color singlet configuration, 
the opposite is true for QCD tree level diagrams.
These considerations alone may lead us to believe that QCD 
gives rise to processes with a higher color radiation accompanying the 
two $b$-quark jets; furthermore, the pattern of this radiation is different. 
%
In QCD processes, in fact, the color connection present at LO between the two 
final state partons and the initial state should give rise to a enhanced 
radiation flow in the planes containing each of the two leading jets and 
the z axis: color coherence prescribes the radiation from the incoming 
and outgoing partons to interfere constructively 
in these regions, while the color singlet produced in the
Z decay will emit soft radiation mainly between the two leading jets.

\begin {figure} [h!]
 \centerline{\epsfig{file=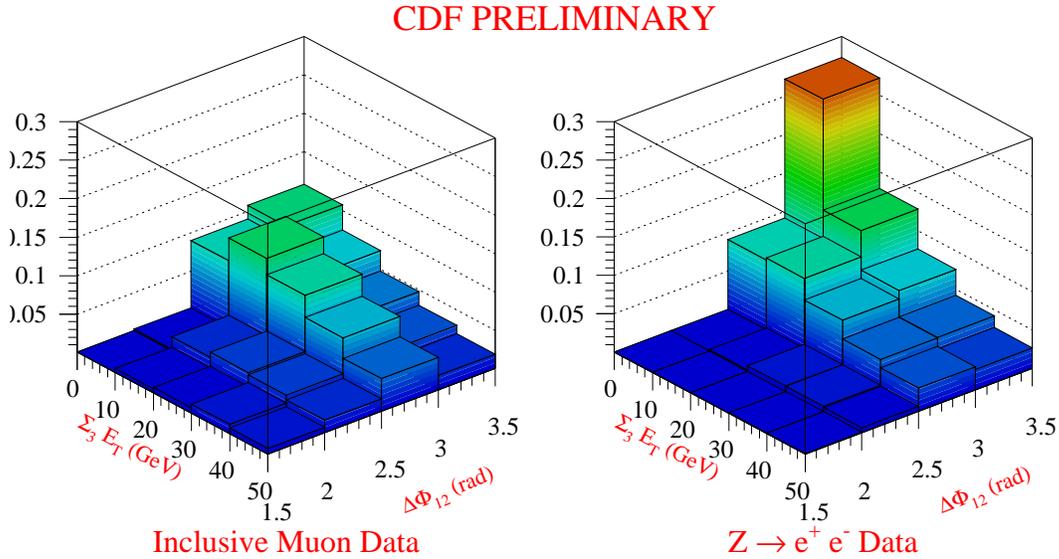,
 bb=25 260 525 550,width=14cm,clip=}} 
 \caption { {\em These plots prove the discriminating power between
            a QCD and a EWK process of the
            two kinematic variables we are selecting our data with. On the
            left is shown our SVX tagged data, on the right the} $Z \to e^+ e^-$
            {\em data. The two distributions are normalized to equal volume. }
            \label{figure:kinvars} }
\end{figure}

Although no event-by-event discrimination appears possible by the use 
of variables that try to pinpoint the differences in the radiation pattern,
two variables that deal inclusively with these features prove
useful for our search. These are the azimuthal 
angle between the two $b$-quark jets, $\Delta \Phi_{12}$, and the sum of 
transverse energies of all the calorimeter clusters in the event
beyond the two jets, $\Sigma_3 E_T$: 
they both have some discriminating 
power between a high radiation process and the colorless production of 
a dijet system; but, while the first one is easy to 
simulate for Monte Carlo programs, being relatively independent from the modeling of 
the underlying event, the second is critically dependent on the detailed 
features of the initial state radiation mechanisms. 
For an homogeneous comparison one is therefore bound to use the experimental
data to understand the behavior of the $\Sigma_3 E_T$. 
The distribution of the SVX data (to be considered a pure background sample,
due to the very low S/N ratio)
in the plane of the two kinematic variables can thus be compared to that
of experimental Z decays to $e^+ e^-$ pairs\footnote {
   These two variables have been shown to be rather insensitive to the
   final state radiation, that is present in the searched process but is
   absent in the leptonic final state of a Z decay. }, 
as shown in fig.\ref{figure:kinvars}. The difference
in the radiation flow for these two samples is evident.
   
\noindent
We therefore select our data by placing tight cuts on these two variables:
we require the two leading jets to be separated in azimuth by at least three
radians, and the additional clusterized $E_T$  be less than 10 GeV. These
cuts will be shown to maximize the expected signal significance in 
Section \ref{section:choice_bin}, where all the ingredients necessary for the
computation are presented.

\section {The Counting Experiment}

\noindent
After the kinematic selection we expect the
dataset to be still rich of background, with a S/N ratio as high as
1/6 for events with $M_{jj}\sim 90 \, GeV/c^2$. Under such circumstances, the
mass distribution can be used to demonstrate the presence of a signal
only if the background shape is very well under control. 
  
   We use events where only one of the two leading jets carries a SVX tag
as a background-enriched sample, and look for a signal as an excess of events 
in the double SVX data; to obtain a background estimate we rely on the 
observation that the probability of finding a secondary vertex
in a jet is independent on the value of the kinematic variables we have
selected our data with. We divide our data into four subsamples: 
events that fall in the kinematic region selected by our cuts 
($\Delta \Phi_{12}>3$ radians, $\Sigma_3 E_T<10 \, GeV$, in
the following referred to as the ``{\em Signal Zone}''), and events that 
fail those requirements, 
from both the double tagged dataset ---``(++)'' in the following--- 
and from events having only one tagged jet ---hereafter ``(+0)'' events.
We then define a tag probability as the ratio between (++) and (+0) events 
outside the Signal Zone, and extrapolate it inside, obtaining
an absolute background prediction for the double tags in the Signal Zone:

\vspace {.2cm}
$
N^{++}_{exp,in} = N^{+0}_{obs,in} \times (N^{++}_{obs,out} / N^{+0}_{obs,out})
$. 
\vspace {.2cm}

\noindent
The procedure just outlined is carried out for each bin of 
invariant mass of the dijet system: by doing that, we obtain a tag 
probability as a function of the dijet mass, from which we evaluate
an absolute background prediction $N^{++}_{exp,in}(M_{jj})$ that can be 
compared to the observed count in each bin of the mass distribution. 
An excess around $90 \, GeV/c^2$ will be bias-free evidence for the signal, 
if away from there the excesses are compatible with zero.

\subsection {The Choice of Cuts \label{section:choice_bin}}

\noindent
In order to extract the highest possible significance from the excess
of events in the dataset we have studied the expected signal significance
as a function of the cuts on the kinematic variables used for the selection.
Our definition of the significance for the present purpose is an  
approximation: we define it as $S=N_{signal} / (\sigma_{bgr}^{stat} \oplus 
\sigma_{bgr}^{syst} \oplus \sqrt{N_{tot}}) $. The Monte Carlo is used to
estimate $N_{signal}$ as a function of the cuts, while $N_{tot}$ is defined as
the sum of $N_{signal}$ and the expected background, computed
with the method described above.
The statistical and systematic errors on the background prediction are added in
quadrature  to the Poisson fluctuation
of the total expected number of events $N_{tot}$, giving a value of S per each
choice of the cut on the variable studied.
We are thus able to decide what are the best possible cuts on the two selection
variables. 

\begin {figure} [h!]
  \begin{minipage}{0.327\linewidth}
    \centerline{\epsfig{file=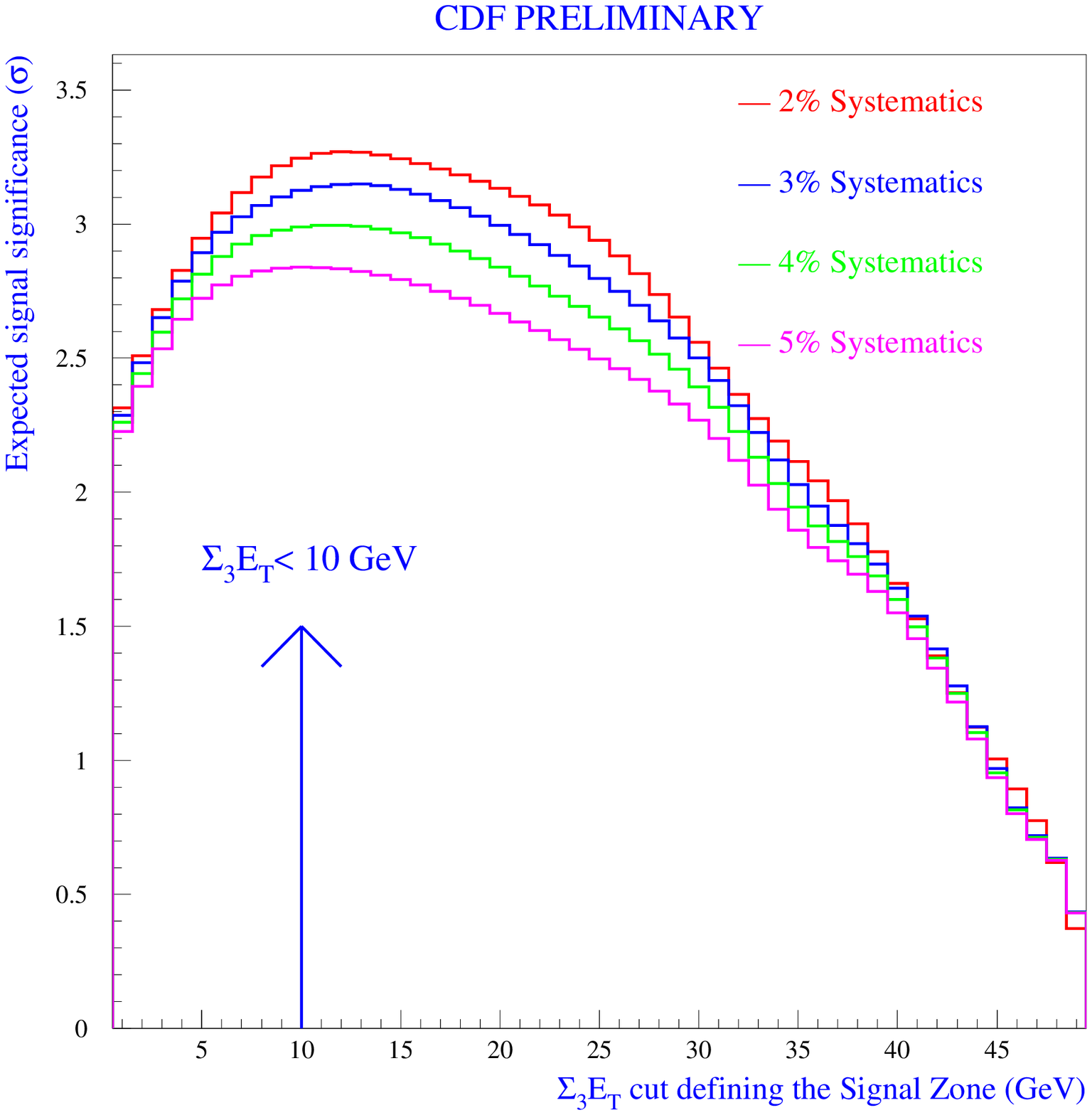,
    bb=35 18 509 535,width=5cm,clip=}}
  \end{minipage}
  \begin{minipage}{0.327\linewidth}
    \centerline{\epsfig{file=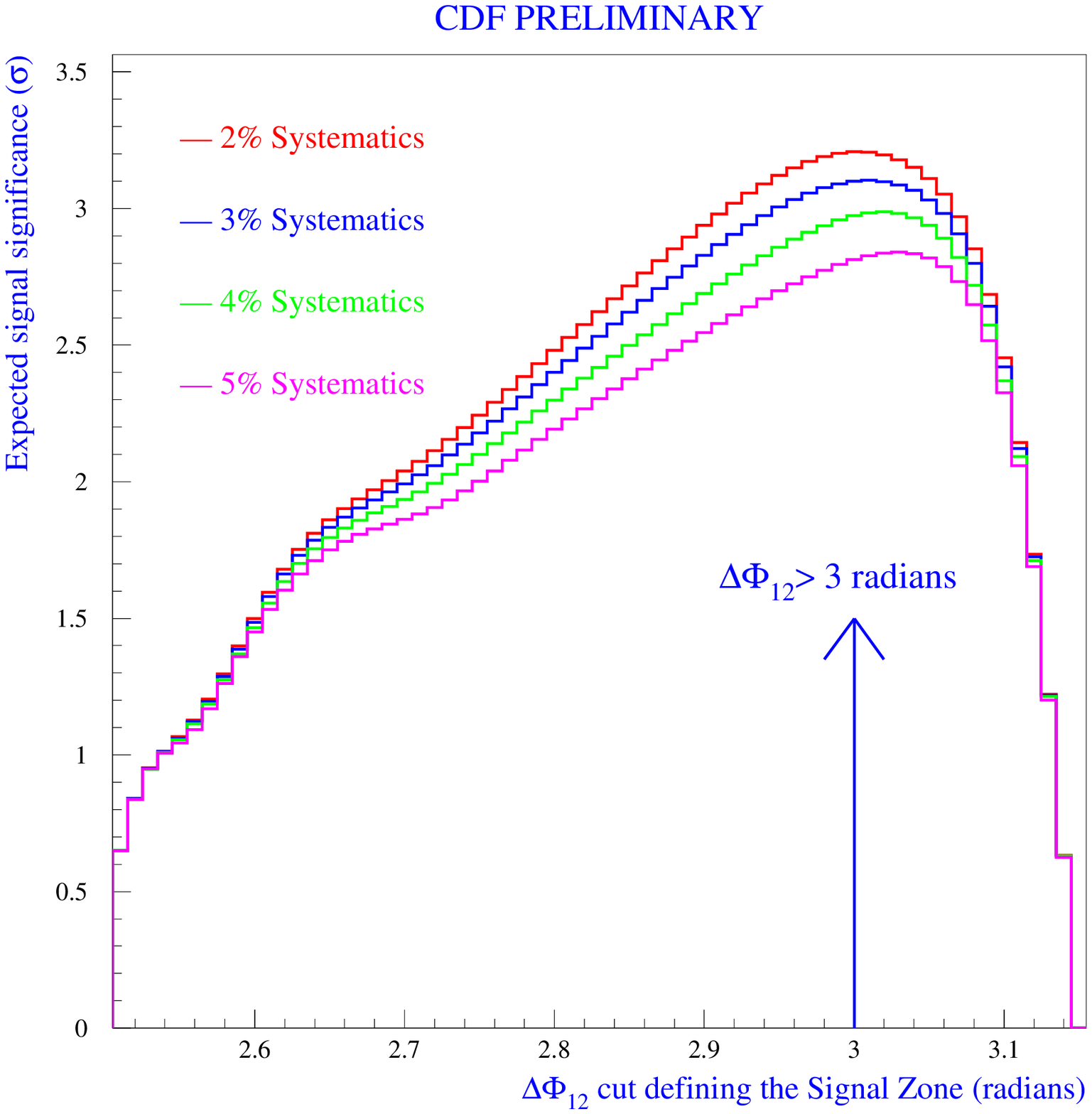,
    bb=35 18 509 535,width=5cm,clip=}}
  \end{minipage}
  \begin{minipage}{0.327\linewidth}
    \centerline{\epsfig{file=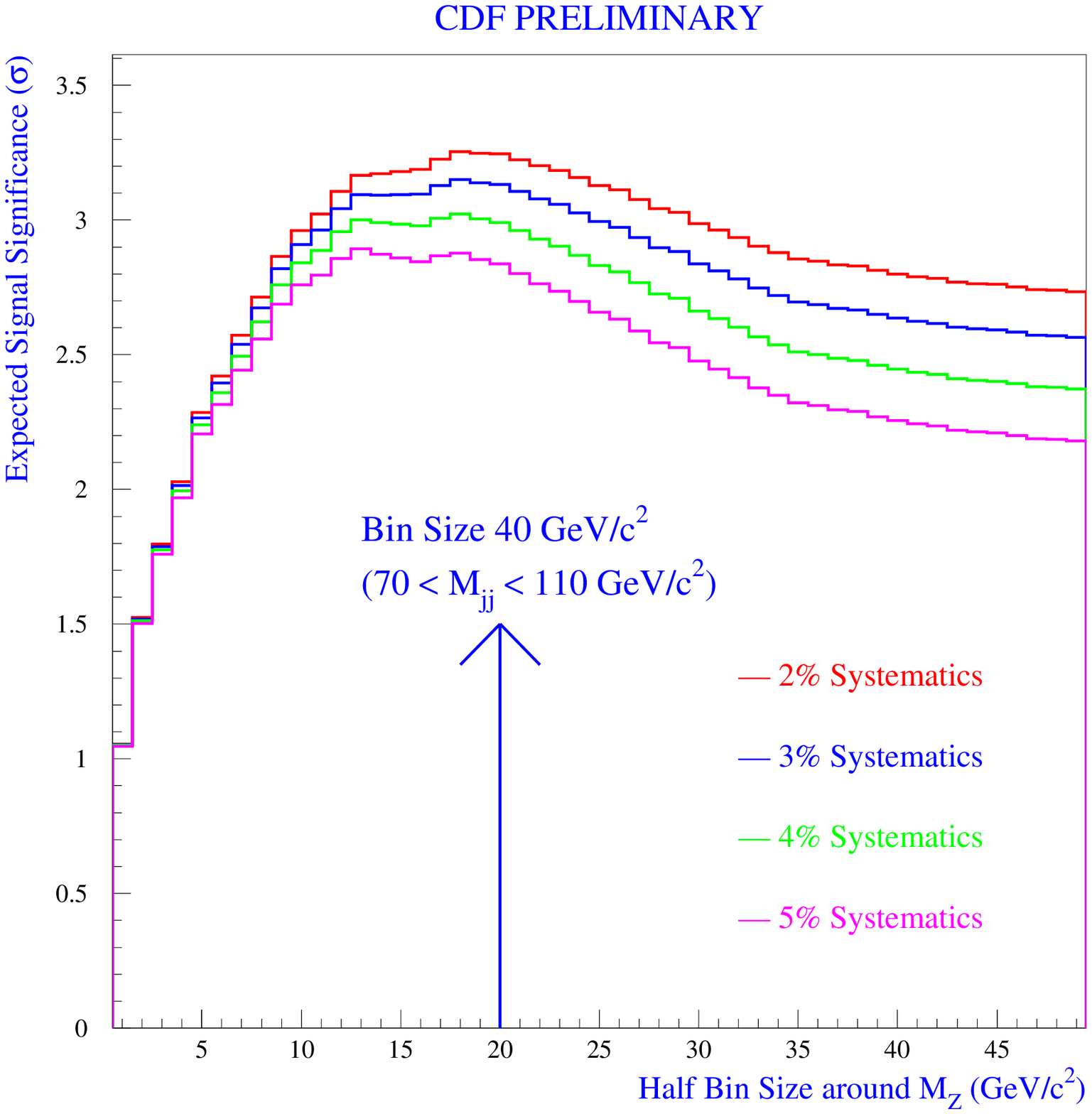,
    bb=35 18 509 535,width=5cm,clip=}}
  \end{minipage}
  \caption { {\em Maximization of the expected signal significance as a function
           of the cut on the kinematic variables. Left: choice of cut on } 
           $\Sigma_3 E_T${\em; center: choice of the cut on } $\Delta \Phi${\em; 
           right: choice of the bin size. Four different possibilities 
           for a systematic uncertainty in the background evaluation are considered. }}
  \label{figure:choice_kin}
\end{figure}

\noindent 
As can be seen in fig.\ref{figure:choice_kin}, the chosen cuts on $\Delta \Phi$
and $\Sigma_3 E_T$ do maximize the expected signal significance S, regardless of
the systematic uncertainty attributed to the extrapolation of the tagging probability.

   The same machinery can be used to decide the optimal binning in the
dijet mass distribution for the counting experiment. 
The dijet mass is of course a very discriminant variable:
by defining appropriately the width of the bin around $90 \, GeV/c^2$ where we look
for an excess of events over the background, we can
again maximize the expected significance. Using the background prediction and the 
expected shape of the signal mass peak from Monte Carlo, we obtain
significance curves that point to the best binning. The latter is chosen to be 
$40 \, GeV/c^2$ wide, as shown in fig.\ref{figure:choice_kin}.

\subsection {Results of the Counting Experiment \label{section:results}}

\noindent
Having tuned cuts and binning to their most favorable values,
we can perform the computation of the background and study the results. 
These are detailed in table \ref{table:count_40}. 

\begin {figure} [h!]
\centerline{\epsfig{file=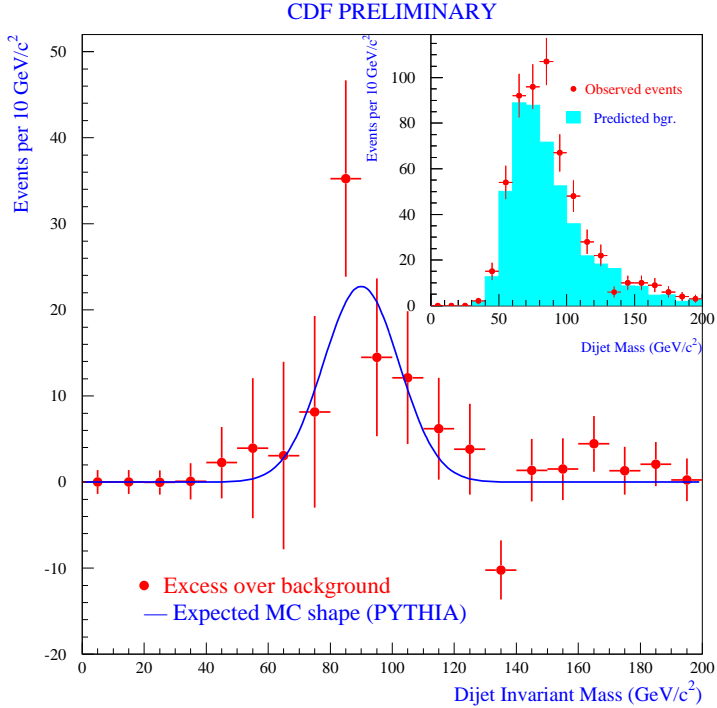,
bb=0 0 550 550,width=10cm,clip=}}
\caption { {\em  Results of the counting experiment with a $10 \,GeV/c^2$ binning. 
         The excess over background
         predictions is compared to the expected shape of a $Z \to b\bar{b}$ 
         mass signal (PYTHIA). The inset shows the mass distribution of
         double tags (points with error bars) and the expected background 
         (full histogram). }}
         \label{figure:count_10}
\end{figure}

\begin{table}[htb]
\begin{center}
\begin{tabular}{||r||c|c|c|c||}
\hline
\hline
Mass Interval & Observed & Expected        & Excess & Exp. Z \\
\hline
$ $           & $ $      & $               $ & $      $ &  \\
$  0- 30 \, GeV/c^2$   &   0      & $   0.05\pm0.09 $ & $-0.05\pm  1.40$ &  $            $ \\ 
$ 30- 70 \, GeV/c^2$   & 163      & $ 149.30\pm6.13 $ & $13.70\pm 14.16$ &  $ 1.7\pm  0.8$ \\
$ 70-110 \, GeV/c^2$   & 318      & $ 248.49\pm8.95 $ & $69.51\pm 19.95$ &  $41.4\pm 18.8$ \\
$110-150 \, GeV/c^2$   &  66      & $  65.51\pm4.68 $ & $ 0.49\pm  9.38$ &  $ 2.4\pm  1.1$ \\
$150-190 \, GeV/c^2$   &  29      & $  19.70\pm2.59 $ & $ 9.30\pm  5.97$ &  $            $ \\
$190-230 \, GeV/c^2$   &   7      & $   7.06\pm1.73 $ & $-0.06\pm  3.16$ &  $            $ \\
$230-270 \, GeV/c^2$   &   3      & $   2.24\pm0.82 $ & $ 0.78\pm  2.38$ &  $            $ \\
$270-310 \, GeV/c^2$   &   1      & $   1.60\pm0.95 $ & $-0.60\pm  2.00$ &  $            $ \\
$310-350 \, GeV/c^2$   &   0      & $   0.55\pm0.98 $ & $-0.55\pm  1.71$ &  $            $ \\
$350-390 \, GeV/c^2$   &   1      & $   0.00\pm0.00 $ & $ 1.00\pm  1.76$ &  $            $ \\
$ $           & $ $      & $ $               & $              $ &  $            $ \\
\hline
\hline
\end{tabular}
\caption{ \label{table:count_40}
          {\em Results of the counting experiment.} }
\end{center}
\end {table}
\begin {figure} [h!]
   \begin{minipage}{0.49\linewidth} 
     \centerline{\epsfig{file=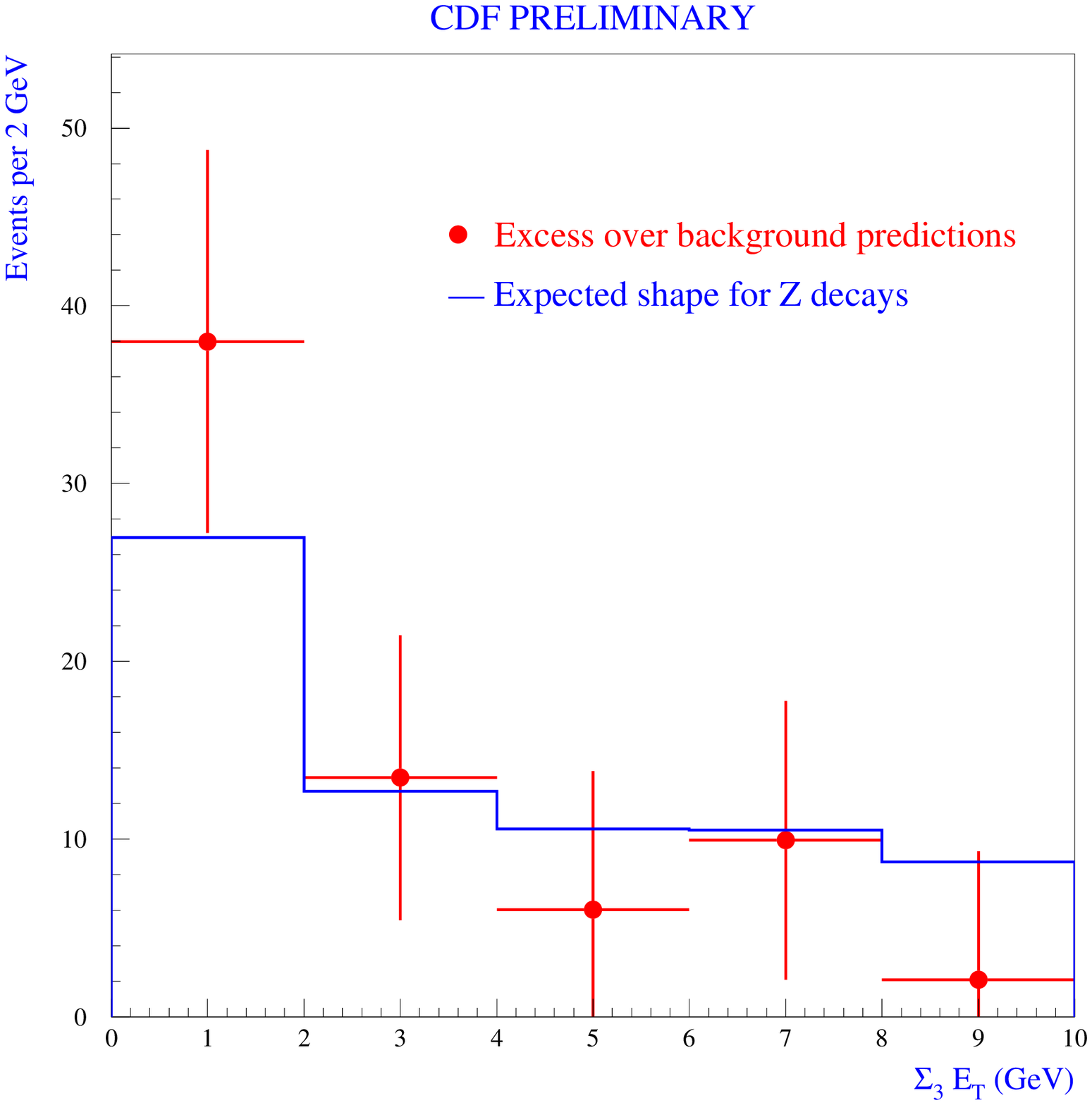,
     bb=5 15 510 535,width=7cm,clip=}}
   \end{minipage}
   \begin{minipage}{0.49\linewidth}
     \centerline{\epsfig{file=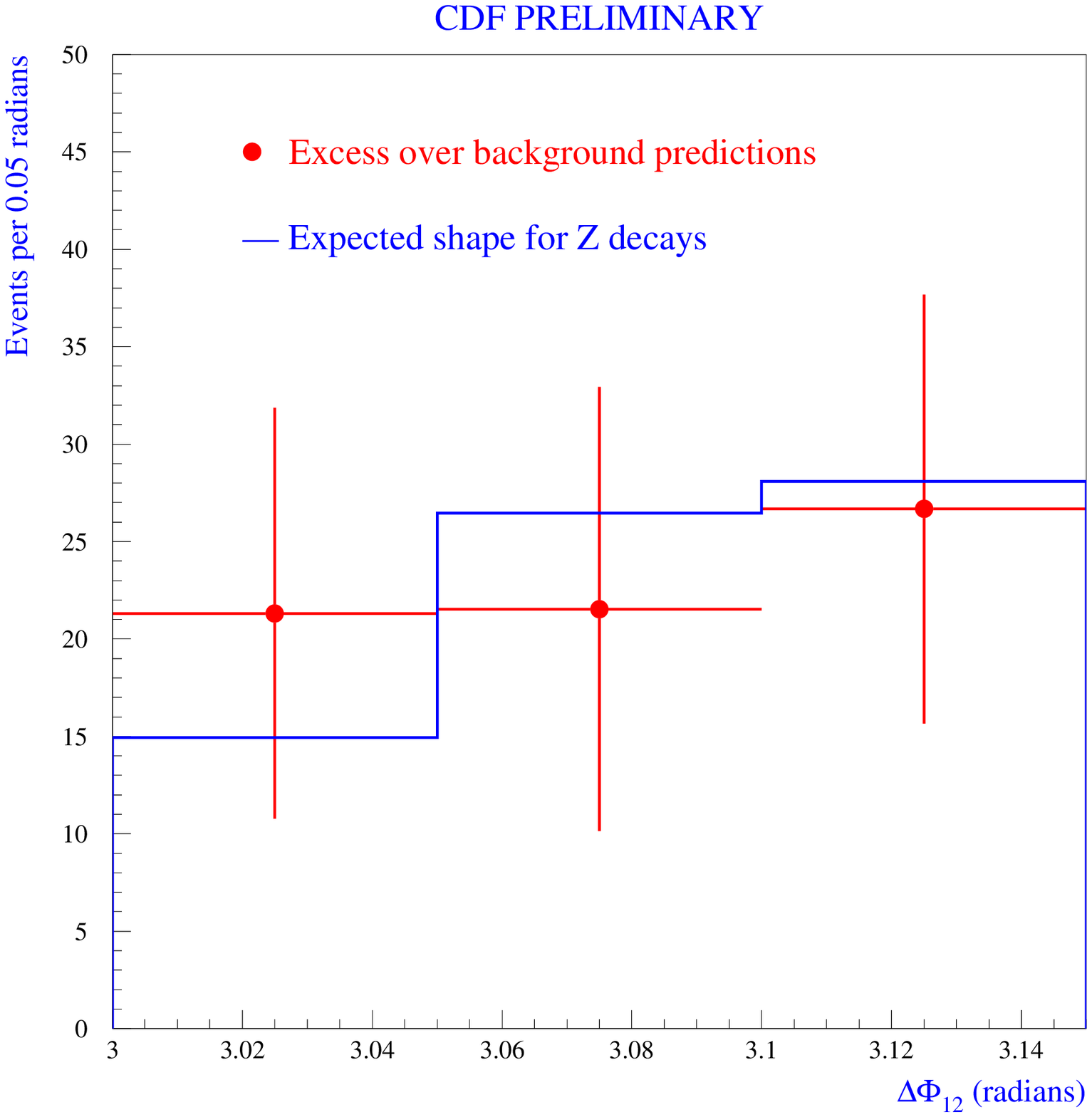,
     bb=5 15 510 535,width=7cm,clip=}} 
   \end{minipage}
\caption { {\em The excess observed in the} $70 \div 110 \, GeV$ {\em bin is 
           distributed according to expectations in the} $\Sigma_3 E_T$
           {\em variable (left) and in the} $\Delta \Phi$ {\em variable (right). }}
           \label{figure:count_kin}
\end{figure}

   The excess in the third bin is quite significant: using a conservative 
estimate of the systematic uncertainty in the extrapolation of the tag
probability (a total of 4\% is estimated from comparisons of observed and predicted
events in signal-depleted samples) 
the prediction becomes $N_{exp}=248.49\pm 13.38$ events. 
The probability of a fluctuation of this
number to the 318 observed events is 0.00061, equivalent to 3.23 standard 
deviations for a one-sided gaussian distribution.

   The numbers quoted above are already compelling evidence for the
presence of $Z \to b\bar{b}$ decays in our dataset. But we can
use a smaller bin size to study the {\em shape} of the excess of events in the
mass distribution: we then expect to see a gaussian peak\footnote{
    A tuning of the jet energy corrections, appropriate for $b$-quark jets
    where one of the two undergoes a semileptonic decay to a muon, has been
    performed to increase the evidence for the signal in the mass distribution.
    It allowed us to increase the relative mass resolution $\sigma_M/M_{jj}$ by 50\%
    and to bring the reconstructed dijet mass to the nominal value in Monte Carlo
    events. }
at $90 \,GeV/c^2$, with a r.m.s. of $12.3 \, GeV/c^2$. 
The results of the counting experiment with this smaller bin size confirm
the expectations: the excess fits very well to the 
expected signal shape (fig.\ref{figure:count_10}).

   We have still another method to verify the assumption that we are observing 
a Z signal. In fact, the counting method allows a study of the behavior of 
the excess as a function of the same kinematic variables used for the 
data selection, that have been shown to have a distinctive shape for the electroweak
process. To do that, we select (++) and (+0) events in the
Signal Zone that fall in the interval $70 < M_{jj} < 110 \, GeV/c^2$: as
table \ref{table:count_40} shows, we have an excess of 
$69.5\pm 20.0$ events there. We can then build the $\Sigma_3 E_T$ and
$\Delta \Phi$ distributions
for the (++) events and compare them to the corresponding (+0) distributions 
scaled down by the tag probability 
$P_{90} = N^{++}_{obs,out}(70 \div 110) / N^{+0}_{obs,out} (70 \div 110)$. 
Indeed, the double SVX tags in excess are distributed
in the two kinematic variables as expected for Z events, as can be seen in 
fig.\ref{figure:count_kin}.

\section {Unbinned Likelihood Fits to the Mass Distribution}

Having established the presence of a $Z \to b\bar{b}$ signal in our dataset
with a counting method, we can perform a two-component fit to the dijet mass 
distribution, and extract additional information from its shape. 

\begin{figure}[htb]
\centerline{\epsfig{file=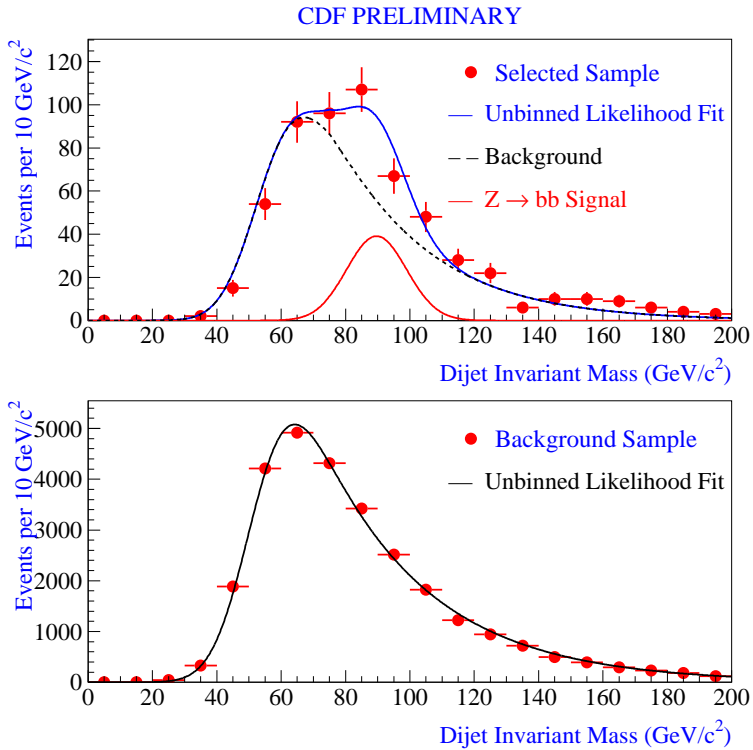,
            bb=0 0 530 530,width=10cm,clip=}} 
\caption { {\em Top: results of the unbinned likelihood fit to the dijet mass 
                distribution in the (++) sample; bottom: results of the unbinned 
                likelihood fit to the dijet mass distribution of (+0)
                events rejected by the kinematic cuts. } } 
\label{figure:likfit}
\end{figure}

   We used a three-step procedure to fit the data. First of all, we obtained
a background shape from a unbinned likelihood fit to a signal-depleted 
sample consisting in (+0) events falling outside of the Signal Zone. 
This sample is expected to be totally dominated by
the QCD background, and can be successfully fit to the following simple 
functional form:\\
\vskip.1cm
$P_{bgr}(M_{jj}) = \frac {1}{\lambda} \cdot e^{-M_{jj}/\lambda} \otimes N(\mu,\sigma)$\\
\vskip.1cm
\noindent
where ``$\otimes$'' is the convolution operator, while $N(\mu,\sigma)$ is the 
Normal distribution. The fit is shown in the lower plot of fig.\ref{figure:likfit}.

   The above form cannot be directly used as a background shape for double SVX events,
since the probability of tagging a second jet is correlated to the invariant mass.
To account for that, we obtained a tag probability curve $P(M_{jj})$ 
by performing a $\chi^2$ fit to the ratio of double and single SVX tags rejected
by the kinematic cuts.     

\noindent
The knowledge of the parameters of the two above fits allows us to write 
a two-component unbinned likelihood for the dijet mass 
distribution of the (++) sample, as follows:
\begin{eqnarray*}
 {\cal L}^{++}   & = & P_{oisson}( \, N^{++}, \,
                               ( n^{++}_{sig} +
                                 n^{++}_{bgr} ) \, ) 
                                 \cdot \nonumber 
  \prod_{i=1}^{N^{++}} 
  \frac{ n^{++}_{sig} \cdot P_{sig}(M_{i}) 
        + n^{++}_{bgr} \cdot P_{bgr}^{\prime}(M_{i}) }
       {n^{++}_{sig} + n^{++}_{bgr} } 
         \nonumber \\
         &&
\end{eqnarray*}
where: $N^{++}$ is the observed number of events in the (++) sample;
$n^{++}_{sig}$ and $n^{++}_{bgr}$ are respectively the number of signal 
and background events in the (++) sample; $P_{bkg}^{\prime}(M)$ is the 
normalized background mass shape of the events in the (++) sample, obtained 
from the background fit to the single SVX data and the tag probability curve;   
and, finally, $P_{sig}(M)$ is a normalized gaussian describing the
signal. 

   By maximizing $\log {\cal L}$ with respect to the number of signal
and background events, and to the mass and width of the signal shape,
but keeping frozen the background shape parameters and the tag probability
curve parameters obtained previously, we obtained $n^{++}_{sig}=91 \pm 30$ events, 
$M_Z = 90.0 \pm 2.4 \, GeV/c^2$, and $\sigma_Z = 9.4 \pm 3.5 \, GeV/c^2$; 
the fit results are shown in the upper plot of fig.\ref{figure:likfit}.
The mass and width of the gaussian peak are in perfect agreement
with Monte Carlo expectations ($M=90.0 \, GeV/c^2$, $\sigma_{M}=12.3 \, GeV/c^2$),
while its normalization is in agreement to the excess obtained in Section
\ref{section:results} but is 
slightly larger than the Monte Carlo predictions ($N_Z=45.5\pm 20.7$ events). 
Although partially correlated to the results of the counting experiment, these
numbers give additional qualitative informations on the behavior of the signal
we have isolated.

   Various studies of the possible systematic uncertainties of the fitting procedure
have been performed. The systematic uncertainty due to the negligence of a signal 
contamination in the background sample was estimated to be $\pm 9$ events; 
the systematics due to the parametrization of the tag probability amount to 
$\pm 16$ events; and, finally, the systematic uncertainty due to the modeling of 
the background shape was showed to amount to $\pm 5$ events. 
Therefore the final result of the unbinned likelihood
fit to the mass distribution is the following: $N_Z = 91 \pm 30 (stat) \pm 19 (syst)$
events.

\section {Conclusions}

\noindent
We have searched 5.5 million inclusive muon events collected by CDF during run 1 
for a $Z \to b\bar{b}$ signal.
The very low signal/noise ratio at trigger level 
(less than $10^{-3}$) 
implies that a really strict selection is required in order to isolate a signal. 
We have designed an optimized selection for that purpose by making use 
of double SECVTX 
tagging plus some kinematic criteria suited to the discrimination of an 
electroweak production from the QCD background.
By these means we select 588 events, 318 of whose have a reconstructed
dijet invariant mass between 70 and 110 $GeV/c^2$, where 
the Z decay is expected to yield $41.4\pm 18.8$ events.
By comparing the observed events with the 
expected background we find evidence of the signal, quantifiable in an 
excess of $69.5 \pm 20.0$ events having a 
suggestive shape in the mass distribution. The excess corresponds to a 
$3.23 \sigma$ fluctuation of the background, in the hypothesis of no signal.
Finally, we have used a unbinned likelihood fit 
to the dijet mass distribution to obtain additional evidence of the 
presence of $91 \pm 30$ (stat) $\pm 19$ (syst)
$Z \to b \bar{b}$ decays in the signal sample. 
 
   We thank the Fermilab staff and the technical staffs of the participating
institutions for their vital contributions. We also thank Michelangelo Mangano
and Mike Seymour for many useful discussions. This work was supported by the U.S.
Department of Energy and National Science Foundation; the Italian Istituto Nazionale di
Fisica Nucleare; the Ministry of Education, Science and Culture of Japan; the Natural
Sciences and Engineering Research Council of Canada; the National Science Council 
of the Republic of China; and the A.P.Sloan Foundation.

\end {document}